\documentclass[aps,pra,notitlepage,twocolumn,showpacs,superscriptaddress]{revtex4-1}
\usepackage{amsfonts}
\usepackage{}
\usepackage{amssymb}  
\usepackage{blindtext}
\usepackage{graphicx}  
\usepackage{dcolumn}   
\usepackage{amsmath,bm,amssymb,latexsym,galois,amsthm}   
\usepackage{dsfont} 
\usepackage{mathrsfs}
\usepackage[all,cmtip]{xy}
\usepackage{bm}
\usepackage{subfig}

\theoremstyle{plain}
\theoremstyle{definition}

\newtheorem*{defn*}{Definition}

\theoremstyle{remark}

\begin{document}

\widetext


\title{Geometric Local Hidden State Model for Some Two-qubit States}
\author{Bai-Chu Yu}
\affiliation{Key Laboratory of Quantum Information, Chinese Academy of Sciences, School of Physics, University of Science and Technology of China, Hefei, Anhui, 230026, P. R. China}
\affiliation{CAS Center For Excellence in Quantum Information and Quantum Physics, University of Science and Technology of China, Hefei, Anhui, 230026, P. R. China}
\author{Zhih-Ahn Jia}
\affiliation{Key Laboratory of Quantum Information, Chinese Academy of Sciences, School of Physics, University of Science and Technology of China, Hefei, Anhui, 230026, P. R. China}
\affiliation{CAS Center For Excellence in Quantum Information and Quantum Physics, University of Science and Technology of China, Hefei, Anhui, 230026, P. R. China}
\author{Yu-Chun Wu}
\email{Email address: wuyuchun@ustc.edu.cn}
\affiliation{Key Laboratory of Quantum Information, Chinese Academy of Sciences, School of Physics, University of Science and Technology of China, Hefei, Anhui, 230026, P. R. China}
\affiliation{CAS Center For Excellence in Quantum Information and Quantum Physics, University of Science and Technology of China, Hefei, Anhui, 230026, P. R. China}

\author{Guang-Can Guo}\affiliation{Key Laboratory of Quantum Information, Chinese Academy of Sciences, School of Physics, University of Science and Technology of China, Hefei, Anhui, 230026, P. R. China}
\affiliation{CAS Center For Excellence in Quantum Information and Quantum Physics, University of Science and Technology of China, Hefei, Anhui, 230026, P. R. China}

\date{\today}

\begin{abstract}
Adopting the geometric description of steering assemblages and local hidden states (LHS) model, we construct the optimal LHS model for some two-qubit states under continuous projective measurements, and obtain a sufficient steering criterion for all two-qubit states. Using the criterion, we show more two-qubit states that are asymmetric in steering scenario under projective measurements. Then we generalize the geometric description into higher dimensional bipartite cases, calculate the steering bound of two-qutrit isotropic states and make discussion on more general cases.
\end{abstract}

\maketitle


\section{introduction}
\indent After quantum nonlocality was introduced by Einstein, Rosen and Podolsky \cite{b1}, the concept of quantum steering was given by Schr\"{o}dinger \cite{b2}. Consider two distant observers, Alice and Bob, sharing a pair of entangled particles, quantum steering describes the phenomenon that the measurement performed by one side changes the state of the other. Recently quantum steering was recognized as a type of quantum correlations intermediate between entanglement and Bell nonlocality \cite{b3,b4}, and it is intrinsically asymmetric, leading to the existence of one-way steering, an interesting phenomenon \cite{b5,b16,b17}. \\
\indent Let Alice be the steering side and Bob be the steered side, that is, Alice would be the one who performs the measurements, then Bob would check if his local system is genuinely influenced by Alice's measurements. Let $\rho$ be the bipartite state held by Alice and Bob, $\mathfrak{M}_{A}$ be the set of measurements Alice is able to perform, $A$ be a measurement in $\mathfrak{M}_{A}$ and $a$ be one of the outcomes of $A$. To make sure that his system is genuinely influenced by the Alice's measurements instead of some preexisting local hidden states (LHS), Bob must exclude the LHS model:
\begin{equation}
\tilde{\rho}_{A}^{a} =\int p(a|A,\lambda)\rho_{\lambda}q(\lambda)d\lambda,
\end{equation}
where $\tilde{\rho}_{A}^{a}=\mathrm{Tr}_{A}(A_{a}\otimes I_{B}\rho)$ is the \emph{unnormalized} conditioned state on Bob's side after Alice obtains outcome $a$ from measurement $A$, $A_{a}$ is the corresponding measurement operator for Alice and $I_{B}$ is the identity for Bob. The set $\{\tilde{\rho}_{A}^{a}\}$ is referred to as a measurement assemblage \cite{b6}. The variable $\lambda$ is distributed with density $q(\lambda)$. The probability distributions in Eq. (1) must satisfy
\begin{align}
\sum_{a}p(a|A,\lambda)=1\nonumber, &\int q(\lambda)d\lambda=1,\\
\int p(a|A,\lambda)q(\lambda)d&\lambda=p(a|A).
\end{align}
A bipartite state is steerable from Alice to Bob if and only if there is no LHS model $\{p(a|A,\lambda),q(\lambda),\rho_{\lambda}\}$ such that both Eqs. (1) and (2) hold for all $a$ and $A\in \mathfrak{M}_{A}$.\\
\indent To bound the set of steerable states under certain measurement set precisely, we must construct the optimal LHS model for the measurement assemblages. Here the optimal LHS model means that if such LHS model do not satisfy Eqs. (1) and (2) for the assemblage, no other LHS model satisfies them \cite{b3,b7}. In this paper, according to a geometric characterization of steering assemblage and LHS model \cite{b7}, a constructive method to obtain optimal LHS model for some two-qubit states are proposed. Moreover, we show that the optimality of the constructed LHS model can be used to obtain more two-qubit states which are asymmetric in steering scenario, and demonstrate one-way steering on them.\\
 \indent The organization of this paper is as follows. After recalling the geometric characterization and the steering criterion \cite{b7} in Sect. \textrm{II}, a specific optimal geometric model for some two-qubit states is obtained in Sect. \textrm{III}. Then in sect. \textrm{IV}, we proposed a practical sufficient steering criterion for two-qubit states. More asymmetric steerable states are obtained in Sect. \textrm{V} and in Sect. \textrm{VI}, the geometric description of steering is generalized into higher dimensional bipartite cases, the steering bound of two-qutrit isotropic states is calculated, and discussion on more general bipartite states is given.\\
\section{the geometric model and steering criterion}
\indent In this section we review the geometric model and steering criterion in Ref. \cite{b7}. To characterize a measurement assemblage, the shrinked Bloch vectors $\bm{s}_{A}^{a}$ of the \emph{unnormalized} conditioned states
\begin{equation}
\tilde{\rho}_{A}^{a}=p(a|A){\rho}_{A}^{a}=\frac{1}{2}[p(a|A)\bm{I}+\bm{s}_{A}^{a}\cdot\bm{\sigma}],
\end{equation}
are put into a unit sphere $\tilde{B}$ which is called probability Bloch sphere.\\
\indent A two-qubit state $\rho$ can be written in Pauli bases as $\rho= \frac{1} {4}\sum_{u,v=0}^{3} G_{uv} \sigma_{u}\otimes\sigma_{v}$, where $G_{uv}$ is the element of real matrix
$G=\begin{pmatrix}  1&\bm{b}^{t}\\ \bm{a}&\bm{T} \end{pmatrix}$ , $\bm{a}$ and $\bm{b}$ are Bloch vectors, $\bm{T}$ is a $3\times 3$ matrix and superscript $t$ means transposition \cite{b8}. When Alice's side is projected onto a pure state $A_{a}=\frac{1}{2}(\bm{I}+\bm{x}_{A}^{a}\cdot\bm{\sigma})$, Bob's conditioned state becomes
\begin{equation}
\tilde{\rho}_{A}^{a}=\frac{1}{4}[(1+\bm{x}_{A}^{a}\cdot\bm{a})\bm{I}+(\bm{b}+\bm{T}^{t}\bm{x}_{A}^{a})\cdot\bm{\sigma}].
\end{equation}
\indent By comparing Eqs. (3) and (4), we can obtain that $p(a|A)=\frac{1}{2}(1+\bm{x}_{A}^{a}\cdot\bm{a})$ and $\bm{s}_{A}^{a}=\frac{1}{2}(\bm{b}+\bm{T}^{t}\bm{x}_{A}^{a})$. The geometric figure Bob obtains in $\tilde{B}$ under projective measurements by Alice is shaped by $\frac{1}{2}\bm{T}^{t}\bm{x}_{A}^{a}$, translated by $\frac{\bm{b}}{2}$ and independent of $\bm{a}$. The shrinked Bloch vectors obtained by POVM are inside the figures, since any POVM operator can be written as a mixture of some projectors.\\
\indent Then a geometric model to characterize the LHS model for the assemblage (satisfying Eqs. (1) and (2)) is proposed. The geometric model $\mathcal{G}$ for a steering figure is a set of nonnegative distributions $\{q(\bm{\xi})$, $p(a|A,\bm{\xi})\}$ satisfying:\\
\noindent(1) The equations
\begin{align}
\sum_{a}p(a|A,&\bm{\xi})=1,\nonumber\\
\int_{N_{C}} q(\bm{\xi})p(a|A,\bm{\xi})d\bm{\xi}=&p(a|A)\int_{N_{C}}q(\bm{\xi})d\bm{\xi}
\end{align}
hold for all $a$ and $A$, where $N_{C}=N\cup \{C\}$ is the combination area of the surface $N$ and the center $C$ of $\tilde{B}$, $\bm{\xi}$ are unit vectors on surface $N$ or zero vector at center $C$. Strictly speaking, the probability of $\bm{\xi}$ at $C$ should be a discrete $p(\bm{0})$, but for convenience we still write it in the integral form, which satisfies $\int_{C}\hat{q}(\bm{0})d\bm{\xi}=\hat{p}(\bm{0})$ and $\int_{C}\hat{p}(a|A,\bm{0})\hat{q}(\bm{0})d\bm{\xi}=\hat{p}(a|A,\bm{0})\hat{p}(\bm{0})$.\\
(2) The equation
\begin{equation}
\bm{s}_{A}^{a}= \int _{N_{C}} p(a|A,\bm{\xi})q(\bm{\xi})\bm{\xi}d\bm{\xi}
\end{equation}
holds for all $a$ and $A$.\\
\indent Using the geometric model, a steering quantity $\mathbb{S}$ is defined, which represents the integral $\int_{N_{C}}q(\bm{\xi})d\bm{\xi}$ for a geometric model $\mathcal{G}$. Usually there are many different g-models $\{\mathcal{G}_{i}\}$ for a steering figure, the optimal geometric model $\mathcal{G}_{o}$ is the one with $\mathbb{S}_{o}=$min$_{i}\{\mathbb{S}_{i}\}$, where $\mathbb{S}_{i}$ is the steering quantity of $\mathcal{G}_{i}$. Quantity $\mathbb{S}_{o}$ can be used to generate a necessary and sufficient steering criterion: a two-qubit state is unsteerable from Alice to Bob if and only if $\mathbb{S}_{o}\leq1$ for Bob, and the LHS model corresponding to $\mathcal{G}_{o}$ is the optimal LHS model \cite{b7}.\\
\section{Optimal geometric models for 3D Bell diagonal states}
\indent In this section we construct the optimal geometric models for T states \cite{b9}, which can be represented in form of $G$ matrix as $G=\begin{pmatrix}  1&\bm{0}\\ \bm{0}&\bm{T} \end{pmatrix}$. Since steerability is unchanged under local unitaries, the steerability of T states can be completely described by the states with diagonal T matrices \cite{b7}. Such states can be written as
$\rho=\frac{1}{4}(\bm{I}\otimes \bm{I}+\sum_{i=1}^{3}T_{ii}\sigma_{i}\otimes\sigma_{i})$. They're also called Bell diagonal states since they can be obtained by convex combinations of Bell states.\\
\indent For Bell diagonal states, the steering figures under projective measurements are central symmetric about the center $C$ of sphere $\tilde{B}$. Let $S_{D}$ denote such figures. $S_{D}$ could be a dot, a segment (1D), an ellipse (2D) and the surface of an ellipsoid (3D), note that the dimemsion here depends on the rank of matrix $\bm{T}$. All of them could be called steering ellipsoids in a general sense \cite{b8}. Werner states \cite{b10} are also a special type of Bell diagonal states of which steering ellipsoids are spheres for both Alice and Bob.\\
\indent Now we are going to focus on the steerability of 3D Bell diagonal states. An optimal geometric model $\mathfrak{G}=\{q_{G}(\bm{\xi}),p_{G}(a|A,\bm{\xi})\}$ will be constructed for 3D $S_{D}$. There are similar results for steering figures with lower dimensions, which we leave in the appendix A. Note that since the Bell diagonal states with dimensions lower than 3 are inside the convex cone of separable states, they can be proved to be separable using partial transpose \cite{b11,b12}. In spite of this, constructing their optimal geometric model is still interesting, and some of the results reveal direct correlations between steerability and the geometry of steering figures (see appendix A).  \\
\indent Let $\bm{n}_{A}^{a}$ denote the outer normal vector of $S_{D}$ corresponding to $\bm{s}_{A}^{a}$ (see Fig 1), and region $R_{A}^{a}$ be the hemisphere consisting of unit vectors $\bm{v}$ satisfying $\bm{v}\cdot\bm{n}_{A}^{a}\geq0$ on $\tilde{B}$. For every 3D $S_{D}$, the conditioned distribution we construct is
\begin{align}
p_{G}(a|A,\bm{\xi})&=\left \{
\begin{aligned}
1, &\quad \bm{\xi}\in R_{A}^{a},\\
1/2, &\quad |\bm{\xi}|=0,\\
0, &\quad otherwise.
\end{aligned}
\right.
\end{align}
\begin{figure}
\centering
\includegraphics[width=1.65in,height=1.7in]{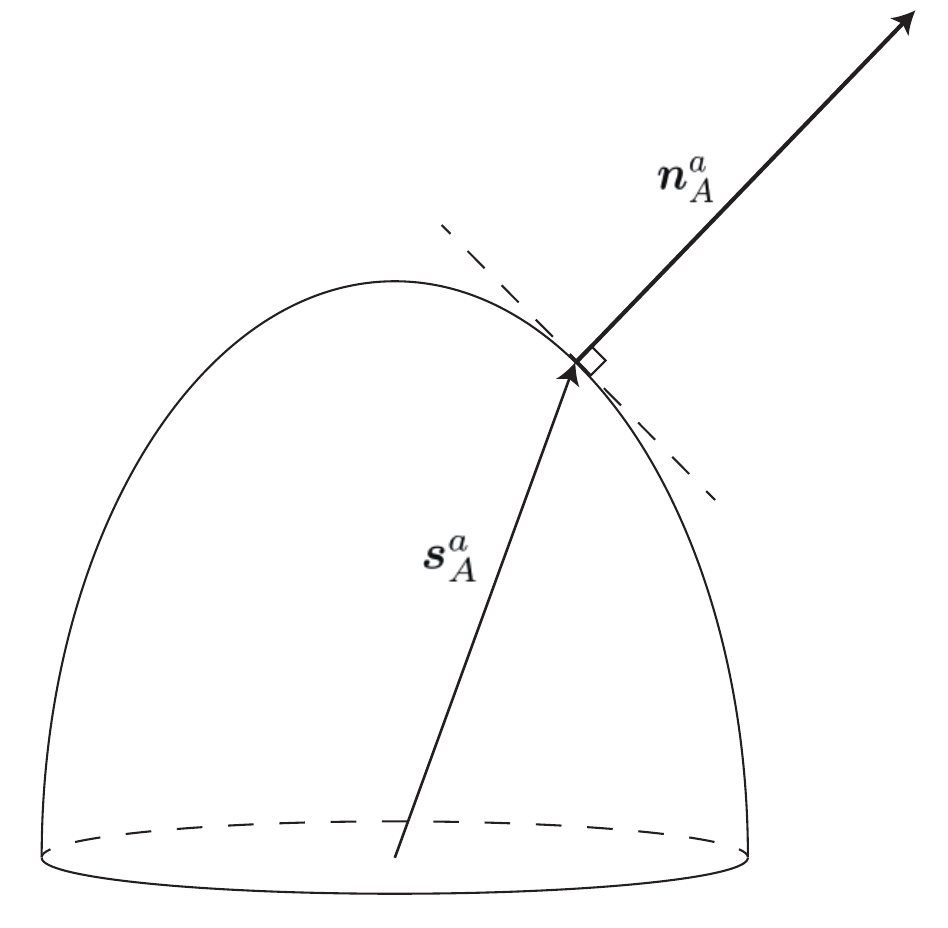}
\caption{The geometric figure is the upper half of an steering figure $S_{D}$. The correspondence of $\bm{s}_{A}^{a}$ and $\bm{n}_{A}^{a}$ is displayed above.}
\label{fig:graph}
\end{figure}

\indent$\bf{Theorem\ 1.}$ If distribution $q_{G}(\bm{\xi})$ exists for a 3D $S_{D}$, model $\mathfrak{G}=\{q_{G}(\bm{\xi}),p_{G}(a|A,\bm{\xi})\}$ would be an optimal geometric model for it.\\
\indent \emph{Proof.} By projecting both sides of equation (6) onto the corresponding $\bm{n}_{A}^{a}$, a new equation is obtained
\begin{equation}
r(\bm{n}_{A}^{a})=\int _{N} p(a|A,\bm{\xi})q(\bm{\xi})\bm{\xi}\cdot\bm{n}_{A}^{a} d\sigma,
\end{equation}
where $r(\bm{n}_{A}^{a})=\bm{s}_{A}^{a}\cdot\bm{n}_{A}^{a}$, $d\sigma$ is the infinitesimal area on surface N. Note that we omit the area $C$ since $\bm{\xi}\cdot\bm{n}_{A}^{a}$ vanishes when $\bm{\xi}=\bm{0}$.\\
\indent Now we add subscripts "$\bm{\xi}_{\pm}$" to outcomes $a$ with respect to each vector $\bm{\xi}$, which indicate the relation between the outcomes $a$ and $\bm{\xi}$. The outcomes $a$ that satisfies $\bm{n}_{A}^{a}\cdot\bm{\xi}\geq0$ are chosen to be $a_{\bm{\xi}+}$, and the others are $a_{\bm{\xi}-}$. Let $c(A,\bm{\xi})$ denote the expression $p(a_{\bm{\xi}+}|A,\bm{\xi})-p(a_{\bm{\xi}-}|A,\bm{\xi})$, where $a_{\bm{\xi}\pm}$ are the two outcomes of measurement $A$. \\
\indent Since all $\bm{n}_{A}^{a}$ are unit vectors, we also place them in the sphere $\tilde{B}$. Then, integrating both sides of Eq. (8) with respect to $\bm{n}_{A}^{a}$ over surface $N$, we have
\begin{equation}
\int_{N}r(\bm{n}_{A}^{a})d\sigma_{\bm{n}}=\int_{N}q(\bm{\xi})\int _{N_{\bm{\xi}}}c(A,\bm{\xi}) \bm{\xi}\cdot\bm{n}_{A}^{a}d\sigma_{\bm{n}}d\sigma_{\bm{\xi}}
\end{equation}
where $N_{\bm{\xi}}$ is the hemisphere consisting of unit vectors $\bm{u}$ satisfying $\bm{u}\cdot\bm{\xi}\geq0$, $d\sigma_{\bm{n}}$ is the infinitesimal area on $N$ corresponding to $\bm{n}_{A}^{a}$. The inner integral with $d\sigma_{\bm{n}}$ is with respect to $\bm{n}_{A}^{a}$ and the outer one with $d\sigma_{\bm{\xi}}$ is with respect to $\bm{\xi}$.\\
\indent For geometric model $\mathfrak{G}$, $c(A,\bm{\xi})=1$. Under $\mathfrak{G}$, Eq. (9) could be simplified as
\begin{equation}
\int_{N}r(\bm{n}_{A}^{a})d\sigma_{\bm{n}}=\int_{N}q_{G}(\bm{\xi})\int _{N_{\bm{\xi}}} \bm{\xi}\cdot\bm{n}_{A}^{a}d\sigma_{\bm{n}}d\sigma_{\bm{\xi}}.
\end{equation}
\indent Let $I_{3}$ denote the integral $\int_{N_{\bm{\xi}}}\bm{\xi}\cdot\bm{n}_{A}^{a}d\sigma_{\bm{n}}$. Its value is $I_{3}=\pi$, independent of $\bm{\xi}$. Then equation (10) becomes
\begin{equation}
\int_{N}r(\bm{n}_{A}^{a})d\sigma_{\bm{n}}=I_{3}\cdot  \int_{N}q_{G}(\bm{\xi})d\sigma_{\bm{\xi}}=\pi\cdot \mathbb{S}_{G}.
\end{equation}
\indent From (11) we obtain
\begin{equation}
\mathbb{S}_{G}=\frac{\int_{N}r(\bm{n}_{A}^{a})d\sigma_{\bm{n}}}{\pi}.
\end{equation}
\indent Consider another geometric model $\mathcal{G}_x$. For any $\mathcal{G}_x$ there is $c_{x}(a|A,\bm{\xi})\leq 1$. Using equation (9) we have
\begin{align}
\int_{N}r(\bm{n}_{A}^{a})d\sigma_{\bm{n}}&=\int_{N}q_{x}(\bm{\xi})\int _{N_{\bm{\xi}}} c_{x}(A,\bm{\xi})\bm{\xi}\cdot\bm{n}_{A}^{a} d\sigma_{\bm{n}} d\sigma_{\bm{\xi}}\nonumber\\ &\leq \int_{N}q_{x}(\bm{\xi})\int _{N_{\bm{\xi}}} \bm{\xi}\cdot\bm{n}_{A}^{a} d\sigma_{\bm{n}} d\sigma_{\bm{\xi}}\nonumber\\
&= \pi \cdot \int_{N}q_{x}(\bm{\xi})d\sigma_{\bm{\xi}}.
\end{align}
Since $\mathbb{S}_{x}=\int_{N_{C}}q_{x}(\bm{\xi})d\bm{\xi}=p(\bm{0})+\int_{N}q_{x}(\bm{\xi})d\sigma_{\bm{\xi}}$, inequality (13) indicates that
$\mathbb{S}_{x}\geq\frac{\int_{N}r(\bm{n}_{A}^{a})d\sigma_{\bm{n}}}{\pi}=\mathbb{S}_{G}$, thus theorem 1 is proved.\qed\\
\indent The existence of $q_{G}(\bm{\xi})$ for 3D Bell diagonal states is showed in some former works \cite{b13,b14,b15}. These works gave a conditioned distribution $p(a|A,\lambda)$ similar to $p_{G}(a|A,\bm{\xi})$, and obtained the expressions for the distribution of $\lambda$, which is corresponding to $q_{G}(\bm{\xi})$ up to a normalizing factor. So we can use Eq. (12) to calculate the optimal steering quantity $\mathbb{S}_{G}$ even without calculating $q_{G}(\bm{\xi})$.\\
\indent Now we calculate the $\mathbb{S}_{G}$ for Werner states as an example. Two-qubit Werner states \cite{b11} can be written as
\begin{equation}
W(p)=p|\psi\rangle \langle\psi|+(1-p)\bm{I}/4
\end{equation}
where $|\psi\rangle \langle\psi|$ is the singlet state and $\bm{I}$ the identity.\\
\indent The steering ellipsoid $S_{D}$ of $W(p)$ is a sphere of radius $p/2$. Distribution $q_{G}(\bm{\xi})$ for $S_{D}$ exists as a constant depending only on $p$. Using Eq. (12) we obtain that $\mathbb{S}_{G}=2p$ for $W(p)$, so states $W(p)$ admit an LHS model when $p\leq1/2.$\\

\section{Obtaining a sufficient steering criterion under projective measurements}
Using quantity $\mathbb{S}_{G}$, a sufficient steering
criterion for more two-qubit states under projective measurements can be obtained. Earlier we showed that for a two-qubit state
\begin{equation}
\rho=\frac{1}{4}(\bm{I}+\bm{a}\cdot\bm{\sigma}\otimes\bm{I}_{B}+\bm{I}_{A}\otimes\bm{\sigma}\cdot\bm{b}+\sum_{u,v=2}^{4}T_{uv}\sigma_{u}\otimes\sigma_{v}),
\end{equation}
which can also be represented by a coefficient matrix $G=\begin{pmatrix}  1&\bm{b}^{t}\\ \bm{a}&\bm{T} \end{pmatrix}$, its steering ellipsoid is shaped by matrix $\bm{T}$ and translated by $\frac{\bm{b}}{2}$. We call all the ellipsoids which are the same up to some translations and rotations in $\tilde{B}$ congruent ellipsoids, and we call the Bell diagonal state with ellipsoid $S_{D}$ the basic state of all states whose steering ellipsoids are congruent to $S_{D}$.  \\
\indent $\bf{Lemma\ 1.}$  $\mathcal{G}_{i}$ is an arbitrary geometric model for the steering ellipsoid $S$ of a two-qubit states under projective measurements, with a steering quantity $\mathbb{S}_{i}$. The steering ellipsoid of its basic state is denoted as $S_{D}$, with a $\mathbb{S}_{G}$ calculable by former method. Now we have:  $\mathbb{S}_{i}\geq \mathbb{S}_{G}$.\\
\indent \emph{Proof.---} Let $\{q_{i}(\bm{\xi}),p_{i}(a|A,\bm{\xi})\}$ denote geometric model $\mathcal{G}_{i}$. For 3D ellipsoids $S$, the proof of theorem 1 can be directly used for Lemma 1, by substituting $S$ and $\mathcal{G}_{i}$ into the both sides of Eq. (9), we can also obtain the same result in equations and inequations (13) for $q_{i}(\bm{\xi})$, so we have $\mathbb{S}_{i}\geq \mathbb{S}_{G}$. Note that the integral $\int_{N}r(\bm{n}_{A}^{a})d\sigma_{\bm{n}}$ depends only on the shape and the size of the steering ellipsoid, it keeps unchange upon translations of the steering figures, even when some $r(\bm{n}_{A}^{a})$ becomes negative. The cases for lower dimensions are left in appendix B. \qed\\

\indent Lemma 1 shows that $\mathbb{S}_{G}$ for $S_{D}$ is a lower bound of quantity $\mathbb{S}$ for all the congruent ellipsoids of $S_{D}$. Using lemma 1 a criterion for steering can be directly obtained.\\
\indent $\bf{Criterion\ 1.}$ A two-qubit state $W$ is steerable for both directions if $\mathbb{S}_{G}>1$ for the steering figure of its basic state.\\
\section{Demonstration of asymmetric steering}
\indent In Ref. \cite{b5}, a state which exhibits asymmetric one-way steering under all projective measurements was proposed as
\begin{equation}
\rho=\frac{1}{2}\psi_{-}+\frac{1}{5}|0\rangle\langle0|\otimes\frac{\bm{I}_{B}}{2}+\frac{3}{10}\cdot\frac{\bm{I}_{A}}{2}\otimes|1\rangle\langle1|,
\end{equation}
where $\psi_{-}$ is the density matrix of the singlet state $(|0, 1\rangle-|1, 0\rangle)/\sqrt{2}$. State $\rho$ is unsteerable from Bob to Alice but steerable from Alice to Bob. Using geometric models and the results above we can also demonstrate that states
\begin{equation}
\varphi(p)=\frac{1}{2}\psi_{-}+p\phi\otimes\frac{\bm{I}_{B}}{2}+\frac{3}{2}p\cdot\frac{\bm{I}_{A}}{2}\otimes\phi^{\bot}+(\frac{1-5p}{2})\frac{\bm{I}}{4}
\end{equation}
$(0<p\leq\frac{1}{5})$ exhibits asymmetric steering under projective measurements, where $\phi$ is a qubit pure state and $\phi^{\bot}$ is the pure state orthogonal to $\phi$, $\bm{I}$ is the two-qubit identity. Before giving the detailed demonstration, we state that we always let $q(\bm{0})$ vanish in all g-models hereinafter, thus vector $\bm{\xi}$ denote unit vectors only. This would simplify the process without influencing the result.\\
\indent The steering ellipsoids of $\varphi(p)$ are spheres with radius $\frac{1}{4}$, which are congruent to the steering figure of Werner state $W(\frac{1}{2})$ (we denote it as $S_{W}$). The value $\mathbb{S}_{G}$ for $S_{W}$ is 1, from lemma 1 we know that any g-models $\mathcal{G}_{i}$ for $\varphi(p)$ would have quantity $\mathbb{S}_{i}\geq1$.\\
 \indent Let $\bm{u}_{X}$ denote the Bloch vector of $\phi$ and $\bm{u}_{Y}$ denote the one of $\phi^{\bot}$, we have $\bm{u}_{Y}=-\bm{u}_{X}$. Let $S_{Y}$ denote the steering ellipsoid for Bob under projective measurements by Alice and $S_{X}$ denote the ellipsoid for Alice under projective measurements by Bob. Using former results we know that $S_{X}$ is a sphere with radius $\frac{1}{4}$, translated by $\frac{p}{2}\cdot\bm{u}_{X}$, and
 \begin{equation}
 p(b|B)=\frac{1}{2}+\frac{3}{4}p\cos\beta,
 \end{equation}
 where $p(b|B)$ is the probability that Bob gets outcome $b$ under projective measurement $B$, $\beta$ is the angle between $\bm{u}_{Y}$ and the Bloch vector $\bm{x}_{B}^{b}$ of projector $B_{b}$.\\
 \indent \emph{1.Unsteerability from Bob to Alice.}\\
\indent For $S_{X}$, we propose a model $\{q_{X}(\bm{\xi}),p_{X}(b|B,\bm{\xi})\}$ (we'll denote it with $\mathcal{G}_{X}$)
\begin{align}
p_{X}&(b|B,\bm{\xi})=\left \{
\begin{aligned}
1,    &\qquad \bm{\xi}\cdot\bm{n}_{B}^{b}\geq0,\\
0,    &\qquad otherwise,
\end{aligned}
\right.\nonumber\\
&q_{X}(\bm{\xi})=\frac{1}{4\pi}(1+3p\cos\beta'),
\end{align}
where $\beta'$ is the angle between $\bm{\xi}$ and $\bm{u}_{X}$, $\bm{n}_{B}^{b}$ is the outer normal vector of $S_{X}$ at $\bm{s}_{B}^{b}$, $\bm{s}_{B}^{b}$ are the shrinked Bloch vectors that constitute $S_{X}$, corresponding to the assemblage $\{\tilde{\rho}_{B}^{b}\}$. \\
\indent Note that $p_{X}(b|B,\bm{\xi})$ is actually the same as $p_{G}(b|B,\bm{\xi})$ of the basic Bell diagonal state of $\varphi(p)$, which means we just change distribution $q_{G}(\bm{\xi})$ into $q_{X}(\bm{\xi})$, then we obtain above model for $S_{X}$ from $\mathfrak{G}$. Actually, for the geometric model $\mathfrak{G}$ of an arbitrary $S_{W}$, any distributions $\{q'(\bm{\xi}),p'(a|A,\bm{\xi})\}$ that satisfy
\begin{align}
&p'(a|A,\bm{\xi})=p_{G}(a|A,\bm{\xi})\nonumber,\\
q'(\bm{\xi})&+q'(-\bm{\xi})=q_{G}(\bm{\xi})+q_{G}(-\bm{\xi}),
\end{align}
under the same set of projectors $\{A_{a}\}$ would have
\begin{equation}
 \bm{s}_{A}^{'a}=\bm{s}_{W_{A}}^{a}+\bm{t},
 \end{equation}
where $\bm{s}_{A}^{'a}$ and $\bm{s}_{W_{A}}^{a}$ are vectors obtained by substituting $\{q'(\bm{\xi}),p'(a|A,\bm{\xi})\}$ and $\{q_{G}(\bm{\xi}),p_{G}(a|A,\bm{\xi})\}$ into Eq. (6) respectively. Equation (21) shows that the steering figure obtained by $\{q'(\bm{\xi}),p'(a|A,\bm{\xi})\}$ is a congruent ellipsoid of $S_{W}$ with a translation $\bm{t}$. We can see that $\mathcal{G}_{X}$ is such a model, the steering figure it generates by Eq. (6) is a translated sphere with radius $\frac{1}{4}$, and its translation vector $\bm{t}_{X}$ is
\begin{equation}
\bm{t}_{X}=\int_{R_{\bm{u}X}^{+}}\bm{\xi}\cdot\frac{3}{4\pi}p\cos\beta'd\sigma=\frac{p}{2}\cdot\bm{u}_{X},
\end{equation}
where $R_{\bm{u}X}^{+}$ is the hemisphere consisting of unit vectors $\bm{v}$ satisfying $\bm{v}\cdot\bm{u}_{X}\geq0$. Also we calculate the probability $p(b|B)$ that this model produces according to Eqs. (5),
\begin{equation}
p(b|B)=\int_{R_{B}^{b}}q_{X}(\bm{\xi})d\sigma=\frac{1}{2}+\frac{3}{4\pi}p\int_{R_{B}^{b}}\cos\beta'd\sigma,
\end{equation}
where $R_{B}^{b}$ is the hemisphere consisting of unit vectors $\bm{v}$ satisfying $\bm{v}\cdot\bm{n}_{B}^{b}\geq0$ (remember that $\bm{x}_{B}^{b}$ is the Bloch vector of projector $B_{b}$). Using the method of changing reference frame in \cite{b5}, we obtain that
\begin{equation}
\int_{R_{B}^{b}}\cos\beta'd\sigma=\pi\cos\beta.
\end{equation}
Then we have
\begin{equation}
p(b|B)=\frac{1}{2}+\frac{3}{4}p\cos\beta.
\end{equation}
\indent This means that model $\mathcal{G}_{X}$ is a geometric model for $S_{X}$. Simple calculation shows that value $\mathbb{S}_{X}=\mathbb{S}_{G}=1$, thus the LHS model corresponding to $\mathcal{G}_{X}$ is an LHS model for $\varphi(p)$ from Bob to Alice, and $\varphi(p)$ are unsteerable from Bob to Alice. Using lemma 1 we know that $\mathcal{G}_{X}$ is the optimal geometric model for $S_{X}$, thus the LHS model is also the optimal one.\\
\indent \emph{2.Steerabilility from Alice to Bob.}\\
 \indent Similarly, $S_{Y}$ is a sphere with radius $\frac{1}{4}$, translated by $\frac{3p}{4}\cdot\bm{u}_{Y}$, and
 \begin{equation}
 p(a|A)=\frac{1}{2}+\frac{1}{2}p\cos\alpha,
 \end{equation}
 where $\alpha$ is the angle between $\bm{u}_{X}$ and the Bloch vector $\bm{x}_{A}^{a}$ of projector $A_{a}$.\\
 \indent According to lemma 1, if there is an LHS model for $\rho$ from Alice to Bob, there is a geometric model $\{q_{Y}(\bm{\xi}),p_{Y}(a|A,\bm{\xi})\}$ (we denote as $\mathcal{G}_{y}$) for $S_{Y}$ with $\mathbb{S}_{Y}=\mathbb{S}_{G}=1$. Together with Eqs. (5) and (6), this geometric model must satisfy conditions
 \begin{align}
p_{Y}(a|A,\bm{\xi})&=p_{G}(a|A,\bm{\xi}),\nonumber\\
\int_{N_{C}} q_{Y}(\bm{\xi})p_{Y}(a|&A,\bm{\xi})d\bm{\xi}=p(a|A),
\end{align}
where $\{q_{G}(\bm{\xi}),p_{G}(a|A,\bm{\xi})\}$ is the model $\mathfrak{G}$ for the ellipsoid $S_{W}$ of the Werner state $W(\frac{1}{2})$. Also, the translation vector $\bm{t}_{Y}$ from $S_{W}$ should be $\frac{3p}{4}\cdot\bm{u}_{Y}$, that is, equation
 \begin{align}
\int_{R}q_{Y}(\bm{\xi})\cdot\bm{\xi}d\sigma=\frac{3p}{4}\cdot\bm{u}_{Y}
 \end{align}
should hold, where $R$ is an arbitrary hemisphere on $\tilde{B}$. We propose a $q_{Y}(\bm{\xi})$ of the form
\begin{equation}
q_{Y}(\bm{\xi})=\frac{1}{4\pi}(1+2p\cos\alpha'),
 \end{equation}
 where $\alpha'$ is the angle between $\bm{\xi}$ and $\bm{u}_{Y}$. Similar to former result, this $q_{Y}(\bm{\xi})$ can realize the conditioned probability $p(a|A)$. \\
 \indent However, the translation vector $\bm{t}_{Y}$ under the proposed $q_{Y}(\bm{\xi})$ is
\begin{equation}
\bm{t}_{Y}=\int_{R_{\bm{u}Y}^{+}}\bm{\xi}\cdot \frac{1}{2\pi}p\cos\alpha'd\sigma=\frac{p}{3}\cdot\bm{u}_{Y},
\end{equation}
where $R_{\bm{u}Y}^{+}$ is the hemisphere consisting of unit vectors $\bm{v}$ satisfying $\bm{v}\cdot\bm{u}_{Y}\geq0$. We can see that the $\bm{t}_{Y}$ we calculate is not equal to $\frac{3p}{4}\cdot\bm{u}_{Y}$.\\
\indent In appendix C, we prove that any geometric models $\{q(\bm{\xi}),p(a|A,\bm{\xi})\}$ that satisfy Eqs. (27) and
 \begin{equation}
\int _{N_{C}} p(a|A,\bm{\xi})q(\bm{\xi})\bm{\xi}d\bm{\xi}=\bm{s}_{W_{A}}^{a}+\bm{t}
 \end{equation}
under projective measurements have the same translation vectors $\bm{t}$. This means that there is not a geometric model that satisfies (27) and (28) simultaneously for $S_{Y}$. For any other geometric model of $S_{Y}$, its quantity $\mathbb{S}>1$. Therefore there is not an LHS model for $\varphi(p)$ from Alice to Bob, $\varphi(p)$ are steerable in this direction.\\
\section{Generalization to higher dimensional cases}
\indent We have demonstrated that the geometric picture is very useful in characterizing steering of two-qubit state case, in this section we extend it into higher dimensional bipartite state cases, to obtain more general results for two-qudit $(d\geq2)$ states, also we calculate the steering bound of two-qutrit isotropic states and have further discussions. To construct a geometric model for two-qudit states, we do it step by step similar to the two-qubit case. First we introduce the probability Bloch hypersphere $\tilde{B}$, then we depict the steering figure and find the geometric description of LHS model in $\tilde{B}$.\\
\indent The density matrix of a qudit state $\rho$ can be written in SU(d) generator basis $\{\gamma_{1}, \gamma_{2}, ... , \gamma_{d^{2}-1}\}$ as
\begin{equation}
\rho=\frac{1}{d}(\bm{I}+\bm{s}_{d}\cdot\bm{\gamma}),
\end{equation}
where $\bm{s}_{d}$ is a real $d^{2}-1$ dimensional vector with norm $|\bm{s}_{d}|\leq\sqrt{\frac{(d-1)d}{2}}$ \cite{b18},  $\bm{\gamma}=(\gamma_{1}, \gamma_{2}, ... , \gamma_{d^{2}-1})$. Similar to two-qubit case, a two-qudit state $\rho_{AB}$ can also be represented as
\begin{equation}
\rho_{AB}=\frac{1}{d^{2}}(\bm{I}_{AB}+\bm{a}\cdot\bm{\gamma}\otimes\bm{I}_{B}+\bm{I}_{A}\otimes\bm{\gamma}\cdot\bm{b}+\sum_{u,v=1}^{d^{2}-1}T_{uv}\gamma_{u}\otimes\gamma_{v}),
\end{equation}
when Alice is projected onto pure state $A^{a}=\frac{1}{d}(\bm{I}+\bm{x}_{A}^{a}\cdot\bm{\gamma}$), the unnormalized conditioned state of Bob is
\begin{equation}
\tilde{\rho_{A}^{a}}=\frac{1}{d^2}[(1+\frac{2\bm{a}\cdot\bm{x}_{A}^{a}}{d})\bm{I}+(\bm{b}+\frac{2\bm{T}^{t}\bm{x}_{A}^{a}}{d})\cdot\bm{\gamma}],
\end{equation}
comparing to the form
\begin{equation}
\tilde{\rho_{A}^{a}}=\frac{1}{d}[p(a|A)\bm{I}+\bm{s}_{A}^{a}\cdot\bm{\gamma}],
\end{equation}
we have $p(a|A)=\frac{1}{d}(1+\frac{2}{d}\bm{a}\cdot\bm{x}_{A}^{a})$ and $\bm{s}_{A}^{a}=\frac{1}{d}(\bm{b}+\frac{2}{d}\bm{T}^{t}\bm{x}_{A}^{a})$. Therefore we can introduce a probability Bloch sphere $\tilde{B}(\mathds{R}^{d^{2}-1})$ with radius $|\bm{s}_{d}|$. By putting the shrinked Bloch vectors $\bm{s}_{A}^{a}$ into $\tilde{B}$ we can obtain a steering figure for the measurement assemblage. Analogously we can construct the geometric model $\{p(a|A,\bm{\eta}),q(\bm{\eta})\}$ for qudit cases, satisfying\\
\noindent(1) The equations
\begin{align}
\sum_{a}p(a|A,&\bm{\eta})=1,\nonumber\\
\int_{\Lambda_{C}} q(\bm{\eta})p(a|A,\bm{\eta})d\bm{\eta}=&p(a|A)\int_{\Lambda_{C}}q(\bm{\eta})d\bm{\eta}
\end{align}
hold for all $a$ and $A$, \\
(2) The equation
\begin{equation}
\bm{s}_{A}^{a}= \int _{\Lambda_{C}} p(a|A,\bm{\eta})q(\bm{\eta})\bm{\eta}d\bm{\eta}
\end{equation}
holds for all $a$ and $A$, where $\Lambda_{C}=N\cup \{C\}$ is the combination area of the set $\Lambda$, which consists of Bloch vectors of $d$ dimensional pure states, and the center $C$ of $\tilde{B}$, $\bm{\eta}$ are vectors with modulus $|\bm{s}_{d}|$ in $\Lambda$ or zero vector at center $C$, $d\bm{\eta}$ is the measure of $\bm{\eta}$ in set $\Lambda$. Note that not all vectors with modulus $|\bm{s}_{d}|$ on the surface of $\tilde{B}$ represents quantum states, since the matrices corresponding to some vectors do not satisfy the positive semidefinite condition. $\Lambda$ is just a small region on the surface of $\tilde{B}$ \cite{b19}. The steering quantity for qudit cases is defined as $\mathbb{S}=\int_{\Lambda_{C}}q(\bm{\eta})d\bm{\eta}$, and the criterion "LHS exists for the measurement assemblage if and only if $\mathbb{S}_{o}\leq1$" is valid for any dimensions. The proof is similar to two-qubit case, and is left in Appendix D.\\
\indent Now that the essential elements of the geometric model are built, we are able to construct the specific geometric model for some two-qudit states. We start by considering the simplest highly symmetric case, the two-qutrit isotropic state
\begin{equation}
\rho(p)= p|\phi_{+}\rangle\langle\phi_{+}|+\frac{(1-p)}{9}\bm{I},
\end{equation}
where $|\phi_{+}\rangle=\frac{1}{\sqrt{3}}(|00\rangle+|11\rangle+|22\rangle)$. $\rho(p)$ can also be represented in Gell-Mann matrix basis (GGB) \cite{b18}, a set of SU(3) generator, as
\begin{equation}
\rho(p)=\frac{1}{9}[\bm{I}+\sum_{i=2}^{8}T_{ii}(p)\gamma_{i}\otimes\gamma_{i}].
\end{equation}
We see that matrix $\bm{T}$ for $\rho(p)$ is a diagonal matrix, with elements $T_{ii}(p)=\frac{3}{2}p$ $(i=1,3,4,6,8)$ and $T_{ii}(p)=-\frac{3}{2}p$, $(i=2,5,7)$. Almost all qutrit pure states $|\psi\rangle$ can be represented with 4 variants $\{\theta, \phi, \alpha, \beta\}$ as
\begin{align}
|\psi\rangle=&\cos(\theta)|0\rangle+\sin(\theta)\cos(\phi)e^{i\alpha}|1\rangle\nonumber\\
             &\quad +\sin(\theta)\sin(\phi)e^{i\beta}|2\rangle,
\end{align}
where $\theta\in[0,\frac{\pi}{2})$, $\phi\in[0,\frac{\pi}{2}]$, $0\leq\alpha,\beta\leq2\pi$. The pure states with $\theta=\frac{\pi}{2}$ are omitted in (40) since they are two dimensional states, but this does not influence the calculation we do later since these states are of zero volume compared to all other pure states. We can have a pictorial description of the qutrit pure states in the positive octant of $\mathcal{S}^{2}$ \cite{b19}, as in Fig. 2(a).\\
\begin{figure}[b]
\centering
\subfloat[][]{\label{Fig:a}\includegraphics[width=0.25\textwidth]{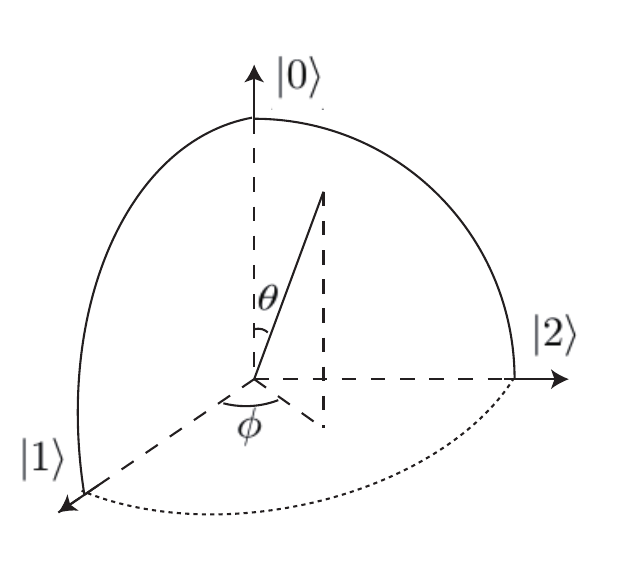}}
	\subfloat[][]{\label{Fig:b}\includegraphics[width=0.25\textwidth]{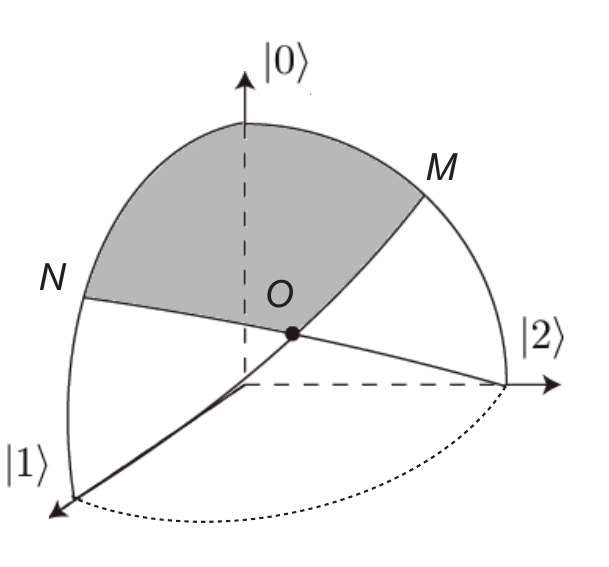}}
\caption{\protect\subref{Fig:a} A pictorial description of almost all pure states. Only $\theta$ and $\phi$ are showed in the figure, each point on the sphere is a torus defined by angles $\alpha$ and $\beta$. The dotted curve on the $\theta=\pi/2$ plane indicates that these states with $\theta=\pi/2$ are omitted. \protect\subref{Fig:b} The grey region is states whose Bloch vectors are in $O_{A}^{a_{0}}$, occupying one third of the total volume of qutrit pure states. }
\label{fig:graph}
\end{figure}
\indent For two-qutrit state $\rho(p)$, when $p>0$ ($p=0$ case is trivial), steering vector $\bm{s}_{A}^{a}$ under projector $A_{a}=|\psi\rangle\langle\psi|$ is proportional to the Bloch vector $\bm{x}'=\frac{2}{3p}\bm{T}^{t}\bm{x}_{A}^{a}$ of another pure state $|\psi\rangle'$, with relation
\begin{equation}
\bm{s}_{A}^{a}=\frac{1}{3}p\cdot\bm{x}'.
\end{equation}
 $|\psi\rangle'$ has the same $\theta$ and $\phi$ with $|\psi\rangle$, but has opposite angles: $-\alpha$ and $-\beta$. From Eq. (41), we also know that steering figure of $\rho(p)$ under projective measurements is similar and proportional to region $\Lambda$, and
\begin{equation}
p=\frac{3|\bm{s}_{A}^{a}|}{|\bm{x}'|}=\sqrt{3}|\bm{s}_{A}^{a}|.
\end{equation}
Therefore, using symmetry we can let $q(\bm{\eta})$ be a uniform distribution, and we just need to choose one projector $A_{a_{0}}=|0\rangle\langle0|$ and build the geometric model for its steering vector $\bm{s}_{A}^{a_{0}}$, the conditioned distribution $p(a|A,\bm{\eta})$ for the other vectors can be obtained using symmetry. The conditioned probability $p(a_{0}|A,\bm{\eta})$ is chosen as
\begin{align}
p(a_{0}|A,\bm{\eta})&=\left \{
\begin{aligned}
1, &\quad \bm{\eta}\in O_{A}^{a_{0}},\\
1/3, &\quad |\bm{\eta}|=0,\\
0, &\quad otherwise,
\end{aligned}
\right.
\end{align}
where $O_{A}^{a_{0}}=\{\bm{u}\in\Lambda| \bm{u}\cdot\bm{s}_{A}^{a_{0}}>\bm{u}\cdot\bm{s}_{A}^{a_{i}}, \forall a_{i}\neq a_{0}\}$. Note that we leave out some vectors in Eq. (43), such as $\{\bm{\eta}|\bm{\eta}\cdot\bm{s}_{A}^{a_{0}}=\bm{\eta}\cdot\bm{s}_{A}^{a_{i}}\}$, but they are of zero volume in total, so it does not affect the result. We depict the region of pure states whose Bloch vectors are in $O_{A}^{a_{0}}$ in Fig. 2(b). With a uniform distribution $q(\bm{\eta})$, model $\{p(a|A,\bm{\eta}),q(\bm{\eta})\}$ satisfies Eqs. (36). Now there remains two conditions to satisfy: Eq. (37) and $\mathbb{S}\leq1$. Since in this case, the components of directions other than that of $\bm{s}_{A}^{a_{0}}$ cancel out in integrating, condition (37) is equivalent to
\begin{equation}
|\bm{s}_{A}^{a_{0}}|= \int_{\Lambda} p(a_{0}|A,\bm{\eta})q(\bm{\eta}) \bm{\eta}\cdot \hat{\bm{a}_{0}} d\bm{\eta},
\end{equation}
where $\hat{\bm{a}}_{0}$ is the unit vector parallel to $\bm{s}_{A}^{a_{0}}$. Using the result in Ref. \cite{b20}, we have
\begin{equation}
d\bm{\eta}= \sin2\phi\cos\theta(\sin\theta)^{3} d\phi d\theta d\alpha d\beta,
\end{equation}
note that the constant factor is omitted. Let $g(\theta,\phi)=\sin2\phi\cos\theta(\sin\theta)^{3}$, combining Eqs. (44) and (45), we have
\begin{align}
|\bm{s}_{A}^{a}|=4\pi^{2}\int_{0}^{\pi/4}\int_{0}^{\pi/2} g(\theta,\phi)\bm{\eta}\cdot \hat{\bm{a}_{0}}q(\bm{\eta})d\phi d\theta\nonumber\\
                +4\pi^{2}\int_{\pi/4}^{\theta_{0}}\int_{Ac(\theta)}^{As(\theta)} g(\theta,\phi)\bm{\eta}\cdot \hat{\bm{a}_{0}}q(\bm{\eta})d\phi d\theta,
\end{align}
where $\theta_{0}=\arccos(\frac{1}{\sqrt{3}})$, $Ac(\theta)=\arccos(\frac{\cos\theta}{\sin\theta})$, $As(\theta)=\arcsin(\frac{\cos\theta}{\sin\theta})$. Using Eqs. (32) and (40) we can obtain
\begin{equation}
\bm{\eta}\cdot \hat{\bm{a}_{0}}=\frac{3^{2}}{2\sqrt{3}}[(\cos\theta)^{2}-\frac{1}{3}],
\end{equation}
 and after substituting Eq. (47) into Eq. (46) we calculate the integral, obtaining $|\bm{s}_{A}^{a}|=\frac{5\sqrt{3}}{36}\pi^{2}\cdot q(\bm{\eta})$. Then we calculate the quantity $\mathbb{S}$ of this model, that is
\begin{align}
\mathbb{S}=4\pi^{2}\int_{0}^{\pi/2}\int_{0}^{\pi/2} g(\theta,\phi)q(\bm{\eta})d\phi d\theta,
\end{align}
which is calculated to be $\pi^{2}q(\bm{\eta})$. As the existence of LHS requires $\mathbb{S}\leq1$, we have $q(\bm{\eta})\leq\frac{1}{\pi^{2}}$, then there is $|\bm{s}_{A}^{a}|\leq \frac{5\sqrt{3}}{36}$, and using Eq. (42) we have $p\leq\frac{5}{12}$. This model is the optimal geometric model for $\rho(p)$ (we will prove later), so result indicates that two-qutrit isotropic states $\rho(p)$ is steerable if and only if $p>\frac{5}{12}$, consistent with the optimal bound obtained by a former work \cite{b4}. Although we just give the two-qutrit example, the optimal steering bound of higher dimensional isotropic states can also be obtained analogously.\\
\indent Now we come to discuss the more general cases of two-qudit T states with full-rank diagonal matrices $\bm{T}$. For these states, probability $p(a|A)=1/d$ under projective measurements, and the steering figures can be characterized by $2(d-1)$ variables, being $2(d-1)$ dimensional regions contained in the Bloch ball of dimensions $d^{2}-1$. As the symmetry reduces, it is difficult to construct the geometric model directly for these states. We try to tackle the problem by directly extending the results in theorem 1 and Eq. (12), and discuss if the extensions are correct. First we propose a geometric model $\{p_{G}(a|A,\bm{\eta}),q_{G}(\bm{\eta})\}$ analogous to Eq. (7), satisfying
\begin{align}
p_{G}(a|A,\bm{\eta})&=\left \{
\begin{aligned}
1, &\quad \bm{\eta}\in L_{A}^{a},\\
1/d, &\quad |\bm{\eta}|=0,\\
0, &\quad otherwise,
\end{aligned}
\right.
\end{align}
where $L_{A}^{a}=\{\bm{u}\in\Lambda| \bm{u}\cdot\bm{n}_{A}^{a}>\bm{u}\cdot\bm{n}_{A}^{a_{i}}, \forall a_{i}\neq a\}$, $\bm{n}_{A}^{a}$ is the outer normal vector corresponding to $\bm{s}_{A}^{a}$. Then we make direct extension of theorem 1 and Eq. (12):\\
\indent \emph{Direct Extension.} Model $\{p_{G}(a|A,\bm{\eta}),q_{G}(\bm{\eta})\}$ exists as the optimal geometric model for the two-qudit T states, with a relation
\begin{equation}
\mathbb{S}_{G}=\frac{\int_{\Omega}r(\bm{n}_{A}^{a})d\bm{n}}{k_{d}},
\end{equation}
where $\Omega$ is the set of vectors $\{\bm{n}_{A}^{a}\}$, $r(\bm{n}_{A}^{a})=\bm{s}_{A}^{a}\cdot\bm{n}_{A}^{a}$, $d\bm{n}$ is the measure of $\bm{n}_{A}^{a}$ in $\Omega$, $k_{d}=\int_{N_{\bm{\eta}}}\bm{\eta}\cdot\bm{n}_{A}^{a}d\bm{n}$ (for an arbitrary $\bm{\eta}$) is a coefficient depends only on dimension $d$, region $N_{\bm{\eta}}$ consists of vectors $\{\bm{n}_{A}^{a}|\bm{\eta}\in L_{A}^{a}\}$.\\
\indent We prove the extension is correct for two-qudit isotropic states in Appendix E. Now we examine the result with the two-qutrit isotropic states $\rho(p)$ we just calculated. For $\rho(p)$, $\bm{n}_{A}^{a}$ is parallel to $\bm{s}_{A}^{a}$, thus $\bm{n}_{A}^{a}$ is one to one proportional to some $\bm{\eta}'$ with relation $\bm{n}_{A}^{a}=\frac{1}{\sqrt{3}}\bm{\eta}'$, and region $\Omega$ is proportional to region $\Lambda$ with a factor $\frac{1}{\sqrt{3}}$, $N_{\bm{\eta}}$ has similar structure as the grey region in Fig. 2(b). Then there are
\begin{align}
\int_{\Omega}r(\bm{n}_{A}^{a})d\bm{n}=\frac{4}{3}\pi^{2}p\int_{0}^{\pi/2}\int_{0}^{\pi/2} g(\theta,\phi)d\phi d\theta,
\end{align}
and
\begin{align}
k_{d}=\frac{4}{\sqrt{3}}\pi^{2}\int_{0}^{\pi/4}\int_{0}^{\pi/2} g(\theta,\phi)f(\theta) d\phi d\theta\nonumber\\
                +\frac{4}{\sqrt{3}}\pi^{2}\int_{\pi/4}^{\theta_{0}}\int_{Ac(\theta)}^{As(\theta)} g(\theta,\phi) f(\theta) d\phi d\theta,
\end{align}
where $f(\theta)$ is the right side of Eq. (47). Calculation shows that $\int_{\Omega}r(\bm{n}_{A}^{a})d\bm{n}=\frac{\pi^{2}}{3}p$, $k_{d}=\frac{5}{36}\pi^{2}$, then we have $\mathbb{S}_{G}=\frac{\int_{\Omega}r(\bm{n}_{A}^{a})d\bm{n}}{k_{d}}=\frac{12}{5}p$, the bound of steerability is $p>\frac{5}{12}$, supporting the former result.\\
 \indent For more general T states, the correctness of such extension is still unknown, since the steering figure (and hence region $\Omega$) may be a $2(d-1)$ dimensional region having different structure with $\Lambda$, and $k_{d}$ might not be independent of $\bm{\eta}$. However, we believe that it is possible to find other T states that fit in the extension. One probable method is to find T states whose steering figure has some symmetry pattern similar to $\Lambda$. More specifically, we can start from T states whose steering figures have symmetry between $d$ steering vectors $\bm{n}_{A}^{a_{i}}$ $(i=0,1,...,d-1)$ of any measurement $A$. We think it is worthwhile doing so since we can get the steering bound of more general states without building specific distribution $q(\bm{\eta})$. What is more, such results may be extended into sufficient steering criterion like criterion 1, or be used to explore the asymmetric steering of higher dimensional bipartite states.\\

 \section{conclusion and discussion}
 \indent We have proposed a specific geometric model $\mathfrak{G}$, and shown that the model is the optimal geometric model for Bell diagonal states. Also we have provided a way to calculate the steerability $\mathbb{S}_{G}$ of $\mathfrak{G}$ without calculating its distribution $q_{G}(\bm{\xi})$. The quantity $\mathbb{S}_{G}$ of ellipsoid $S_{D}$ of Bell diagonal state provides a lower bound of quantity $\mathbb{S}$ for all two-qubit states with steering ellipsoids that are congruent to $S_{D}$. Using this result we obtained a sufficient steering criterion and demonstrated asymmetric steering, obtaining more two-qubit states that are asymmetric in steering under projective measurements. And at the end we generalized the geometric model into higher dimensional bipartite cases, obtaining a steering bound for two-qutrit isotropic states and made some discussion about the steering bound of two-qudit T states.\\
 \indent We have also found several interesting questions for further study.  Finding the optimal geometric model for more two-qubit states is very useful since it not only provides a necessary and sufficient criterion of steering, but also can be used to find more states that demonstrates asymmetric steering. And as we demonstrated that the geometric model can be used in higher dimensional bipartite cases, more steering criteria may be found, and the higher dimensional asymmetric steering may be explored. Also, in appendix A we showed that the steerability of 1D and 2D Bell diagonal states under projective measurements has direct correlation with the geometry of their steering figure. Does such correlation exist in more generalized bipartite states cases? It is a question worth further study.\\
\section*{acknowledgements}
 \indent This work was supported by the National Key Research and Development Program of China (Grant No. 2016YFA0301700) and the Anhui Initiative in Quantum Information Technologies (Grant No. AHY080000).

\section*{appendix a: Optimal geometric model for lower-dimension cases}
\indent Now we introduce the model $\mathfrak{G}$ for lower dimensions cases, followed with some discussions. Note that we always let $q_{G}(\bm{0})=0$ and $p_{G}(a|A,\bm{0})=\frac{1}{2}$ for all cases, therefore the $q_{G}(\bm{\xi})$ and $p_{G}(a|A,\bm{\xi})$ we discuss hereinafter are distributions for unit vectors $|\bm{\xi}|=1$.\\
\indent In 1D ellipsoid case, (\textrm{i}) when $S_{D}$ is a dot at $C$, we let $p_{G}(a|A,\bm{\xi})=\frac{1}{2}$ and $q_{G}(\bm{\xi})=0$. $\mathbb{S}_{G}=0$ for this case, so $\mathfrak{G}$ is the optimal geometric model for $S_{D}$; (\textrm{ii}) when $S_{D}$ is a segment of length $L\leq1$ and symmetric about the center $C$. We use $\bm{s}_{A_{0}}^{a_{+}}$ and $\bm{s}_{A_{0}}^{a_{-}}$ to denote the two opposite vectors with length $\frac{L}{2}$, where $A_{0}$ is the corresponding measurement and $a_{\pm}$ are its two outcomes. Let $p_{G}(a_{\pm}|A_{0},\bm{\xi})=\frac{1}{2}(1+\bm{\xi}\cdot\hat{\bm{s}}_{A_{0}}^{a\pm})$ and $q_{G}(\bm{\xi})=\frac{L}{2}\cdot [\delta(\bm{\xi}-\hat{\bm{s}}_{A_{0}}^{a+})+\delta(\bm{\xi}-\hat{\bm{s}}_{A_{0}}^{a-})]$, where $\hat{\bm{s}}_{A_{0}}^{a\pm}=\frac{\bm{s}_{A_{0}}^{a\pm}}{|\bm{s}_{A_{0}}^{a\pm}|}$, $\delta(\bm{\xi})$ is the Dirac delta function. Such model reproduces $\bm{s}_{A_{0}}^{a_{\pm}}$ while satisfying Eqs. (5) and (6), thus reproducing all other $\bm{s}_{A}^{a}$ on the segment. For $\mathfrak{G}$, quantity $\mathbb{S}_{G}=L$, being equal to the length of the segment. This implies that all Bell diagonal states with an 1D steering ellipsoid are unsteerable. We will prove the optimality of $\mathfrak{G}$ later.\\
\indent For 2D ellipsoid (ellipse) case, we let
\renewcommand{\theequation}{A\arabic{equation}}
\setcounter{equation}{0}
\begin{align}
p_{G}(a|A,\bm{\xi})=&\left \{
\begin{aligned}
1, &\qquad \bm{\xi}\in \tilde{E}_{A}^{a},\\
0, &\qquad \bm{\xi}\in \tilde{E}/\tilde{E}_{A}^{a},\\
\frac{1}{2}, &\qquad \bm{\xi}\notin\tilde{E},
\end{aligned}
\right.
\end{align}
where $\tilde{E}_{A}^{a}$ is the semicircle consisting of the unit vectors $\bm{v}$ satisfying $\bm{v}\cdot\bm{n}_{A}^{a}\geq0$ in $\tilde{E}$, $\bm{n}_{A}^{a}$ is the outer normal vector of the steering ellipse corresponding to $\bm{s}_{A}^{a}$, $\tilde{E}$ is the unit circle on sphere $\tilde{B}$ and in the same plane with the steering ellipse.\\
\indent Suppose valid $q_{G}(\bm{\xi})$ that can reconstruct the ellipse exists and satisfies (\textrm{i}) $q_{G}(\bm{\xi})=0$ when $\bm{\xi}\notin\ \tilde{E}$. This means that $q_{G}(\bm{\xi})$ includes an one-dimensional delta function. Here we omit this delta function and directly treat $q_{G}(\bm{\xi})$ as the 1D distribution on $\tilde{E}$, and $d\bm{\xi}$ becomes $d\theta$ in this case; (\textrm{ii}) $q_{G}(\bm{\xi})$ is central symmetric about center $C$. Then to obtain its specific form, we choose two vectors $\bm{s}_{A}^{a}$, $\bm{s}_{A'}^{a_{'}}$ with a small angle $d\theta$ and get a difference vector $d\bm{s}=\bm{s}_{A}^{a}-\bm{s}_{A'}^{a_{'}}$ by subtraction. Under model $\mathfrak{G}$ we have
\begin{equation}
d\bm{s}=[q_{G}(\bm{\xi}_{s})+q_{G}(\bm{-\xi}_{s})]\bm{\xi}_{s}d\theta',
\end{equation}
where $\bm{\xi}_{s}$ is the unit vector with the same direction of $d\bm{s}$ in $\tilde{E}$, $d\theta'$ is the angle of the non-intersecting parts of semicircles $\tilde{E}_{A}^{a}$ and $\tilde{E}_{A'}^{a'}$, it is also the angle between $\bm{n}_{A}^{a}$ and $\bm{n}_{A'}^{a'}$. To make the case clearer we depict it in FIG.A1.
\begin{figure}
\setcounter{figure}{0}
\centering
\includegraphics[width=1.65in,height=1.7in]{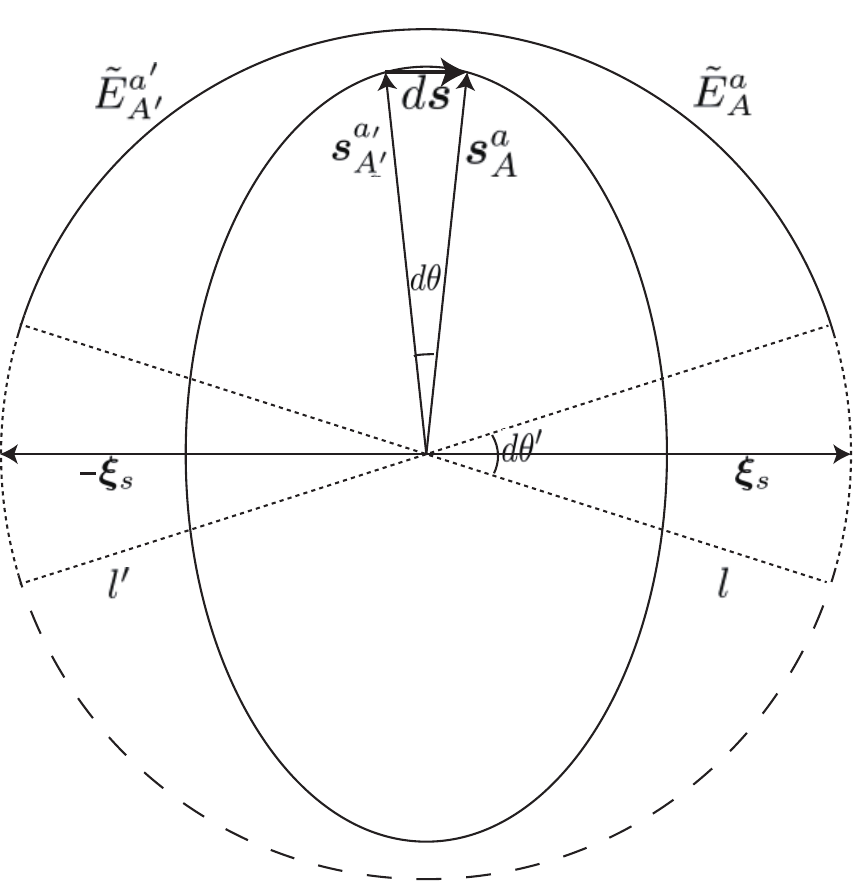}
\renewcommand{\figurename}{FIG.A}\\
\caption{The circle outside is $\tilde{E}$ and the inside figure is the steering ellipse. The bolded vector is $d\bm{s}$, the two dotted parts are the non-intersecting parts of $\tilde{E}_{A}^{a}$ and $\tilde{E}_{A'}^{a'}$, the solid part is the intersecting part, $l\cdot\bm{n}_{A}^{a}=l'\cdot\bm{n}_{A'}^{a'}=0$. }
\label{fig:graph}
\end{figure}
Since $q_{G}(\bm{\xi})$ is central symmetric, we have
\begin{equation}
d\bm{s}=2q_{G}(\bm{\xi}_{s})\bm{\xi}_{s}d\theta'.
\end{equation}
Since
\begin{equation}
|d\bm{s}|=\frac{|\bm{s}_{A}^{a}|}{\cos\alpha}d\theta,
\end{equation}
where $\alpha$ is the angle between $\bm{s}_{A}^{a}$ and $\bm{n}_{A}^{a}$, combining Eqs. (A3) and (A4) we obtain that
\begin{equation}
q_{G}(\bm{\xi}_{s})=\frac{|\bm{s}_{A}^{a}|}{2\cos\alpha}\frac{d\theta}{d\theta'}.
\end{equation}
\indent Equation (A5) shows that calculable $q_{G}(\bm{\xi})$ exists for any 2D $S_{D}$ under $p_{G}(a|A,\bm{\xi})$. However if we just need to calculate the value $\mathbb{S}_{G}$, we don't need to specifically know distribution $q_{G}(\bm{\xi})$. If we take the module of both sides of (A3) and integrate them, we get the circumference of $S_{D}$ on the left side, and $2\cdot \mathbb{S}_{G}$ on the right side. This means that value $\mathbb{S}_{G}$ for any 2D $S_{D}$ equals half of the circumference of the ellipse. After simple calculation, we obtain that $\mathbb{S}_{G}$ for all 2D $S_{D}$ under projective measurements vary from $0.785$ to $1$. This means all 2D Bell diagonal states are unsteerable.\\

\indent $\bf{Theorem\ A1.}$ $\mathfrak{G}$ is the optimal geometric model for 2D and 1D ellipsoids.\\
\indent \emph{Proof.} For 2D case, by projecting both sides of Eq. (6) for 2D ellipsoids onto corresponding $\bm{n}_{A}^{a}$, we get an equation similar to Eq. (8). Integrating both sides of the equation with respect to $\bm{n}_{A}^{a}$ over $\tilde{E}$, we have
\begin{equation}
\int_{\tilde{E}}r(\bm{n}_{A}^{a})d\theta_{\bm{n}}=\int_{N}q(\bm{\xi})\int _{\tilde{E}_{\bm{\xi}}}c(A,\bm{\xi}) \bm{\xi}\cdot\bm{n}_{A}^{a}d\theta_{\bm{n}} d\sigma_{\bm{\xi}},
\end{equation}
where $d\theta_{\bm{n}}$ is the infinitesimal angle corresponding to varying $\bm{n}_{A}^{a}$, $\tilde{E}_{\bm{\xi}}$ is the semicircle generated by the intersection of $N_{\bm{\xi}}$ and circle $\tilde{E}$, $N_{\bm{\xi}}$ is the hemisphere consisting of unit vectors $\bm{u}$ satisfying $\bm{u}\cdot\bm{\xi}\geq0$, $c(A,\bm{\xi})$ is the same as the 3D case. Similar to 3D case, for model $\mathfrak{G}$, $c(A,\bm{\xi})=1$. Then we obtain from Eq. (A6) that
\begin{equation}
\mathbb{S}_{G}=\frac{\int_{\tilde{E}}r(\bm{n}_{A}^{a})d\theta_{\bm{n}}}{I_{2}},
\end{equation}
where $I_{2}$ denotes $\int_{\tilde{E}_{\bm{\xi}}} \bm{\xi}\cdot\bm{n}_{A}^{a}d\theta_{\bm{n}}$ and $I_{2}=2$. For any geometric model $\mathcal{G}_{x}$ of which $c_{x}(A,\bm{\xi})\leq1$, or $q(\bm{\xi})\neq 0$ when $\bm{\xi}\notin \tilde{E}$, the value of the inner integral in Eq. (A6) would be not more than $I_{2}$, thus $\mathbb{S}_{x}\geq \mathbb{S}_{G}$.\qed\\
\indent We can also perform similar procedure for 1D case and get an equation
\begin{equation}
r(\bm{n}_{A}^{a+})+r(\bm{n}_{A}^{a-})=\int_{N}q(\bm{\xi})c(A,\bm{\xi}) \bm{\xi}\cdot\bm{n}_{A}^{a} d\sigma_{\bm{\xi}},
\end{equation}
where $\bm{n}_{A}^{a\pm}$ are two opposite unit vectors parallel to the steering segments, $r(\bm{n}_{A}^{a\pm})=\frac{L}{2}$. Using Eq. (A8) we can prove in a similar way that the model $\mathfrak{G}$ we proposed earlier for 1D case is the optimal one.\\
\indent Also we can summarize an equation for $\mathbb{S}_{G}$ in all cases
\begin{equation}
\mathbb{S}_{G}=\frac{V_{d}}{I_{d}},
\end{equation}
where $d$ represents the dimension of the steering ellipsoid. For 1D and 2D cases, $V_{d}$ equals to the length and circumference of the steering ellipsoids respectively.\\

 \indent Note that the optimal quantity $\mathbb{S}_{o}=\mathbb{S}_{G}$ for a 2D $S_{D}$ is $\frac{1}{2}\mathcal{T}(a+b)$, where $a$ and $b$ are the length of the two semi-axes of the ellipse, $\mathcal{T}$ is the elliptic coefficient. When $b\rightarrow0$, the 2D $S_{D}$ becomes an 1D $S_{D}$. And at the same time, $\mathcal{T}\rightarrow4$, the quantity $\mathbb{S}_{o}\rightarrow2a=L$. This shows that $\frac{1}{2}\mathcal{T}(a+b)$ can be the common expression of quantity $\mathbb{S}_{o}$ for 1D and 2D cases. Then, we may wonder if the quantity $\mathbb{S}_{o}$ of all $S_{D}$ can be written as $\mathcal{K}(a+b+c)$, where $\mathcal{K}$ is a coefficient depends only on the shape, but not the size of the ellipsoid $S_{D}$. This question is left for further study.  \\
\section*{appendix b: Proof of lemma 1 for lower-dimension cases}
\renewcommand{\theequation}{B\arabic{equation}}
\setcounter{equation}{0}
\indent For 2D ellipsoids, the proof of theorem A1 can be directly used for congruent ellipses in the planes that contain the center $C$. For those congruent ellipses in the planes that do not contain $C$ (we denote these planes with $\overline{P}$), we project center $C$, vectors $\bm{\xi}$ and $\bm{s}_{A}^{a}$ onto $\overline{P}$, and denote their projections with $\overline{C}$, $\overline{\bm{\xi}}$ and $\overline{\bm{s}_{A}^{a}}$ respectively. The unit circle centered at $\overline{C}$ is denoted with $\overline{E}$. An projected equation of Eq. (6) can also be obtained as
\begin{equation}
\overline{\bm{s}_{A}^{a}}= \int _{N_{C}} p_{i}(a|A,\bm{\xi})q_{i}(\bm{\xi})\overline{\bm{\xi}}d\sigma.
\end{equation}
Then we obtain the outer normal vectors $\overline{\bm{n}_{A}^{a}}$ of the ellipses on plane $\overline{P}$. By projecting Eq. (B1) on corresponding $\overline{\bm{n}_{A}^{a}}$ and integrating the projected equation with respect to $\overline{\bm{n}_{A}^{a}}$, we can obtain an equation on $\overline{P}$
\begin{equation}
\int_{\overline{E}}\overline{r(\bm{n}_{A}^{a})}d\overline{\theta_{\bm{n}}}=\int_{N}q_{i}(\bm{\xi})\int _{\overline{E_{\xi}}}c_{i}(A,\bm{\xi}) \overline{\bm{\xi}}\cdot\overline{\bm{n}_{A}^{a}}d\overline{\theta_{\bm{n}}} d\sigma_{\bm{\xi}},
\end{equation}
where lines over the terms indicate that they are in plane $\overline{P}$. $\overline{r(\bm{n}_{A}^{a})}=\overline{\bm{n}_{A}^{a}}\cdot\overline{\bm{s}_{A}^{a}}$, $\overline{E_{\xi}}$ is the semicircle consisting of unit vectors $\overline{\bm{u}}$, satisfying $\overline{\bm{u}}\cdot\overline{\bm{\xi}}\geq0$, $d\overline{\theta_{\bm{n}}}$ are infinitesimal angles corresponding to $\overline{\bm{n}_{A}^{a}}$. The left side of (B2) equals $\int_{\tilde{E}}r(\bm{n}_{A}^{a})d\theta_{\bm{n}}$ of the $S_{D}$ which is congruent to $S$. Then we let $q'_{i}(\bm{\xi})=q_{i}(\bm{\xi})\cdot|\overline{\bm{\xi}}|$, $\overline{\bm{\xi}}'=\overline{\bm{\xi}}/|\overline{\bm{\xi}}|$, and for the right side we have
\begin{align}
\int_{N}q'_{i}(\bm{\xi})\int _{\overline{E_{\xi}}}c_{i}(A,\bm{\xi}) \overline{\bm{\xi}}'\cdot\overline{\bm{n}_{A}^{a}}d\overline{\theta_{\bm{n}}} d\sigma_{\bm{\xi}}\nonumber\\
\leq \int_{N}q'_{i}(\bm{\xi})d\sigma_{\bm{\xi}}\cdot\int _{\overline{E_{\xi}}}\overline{\bm{\xi}}'\cdot\overline{\bm{n}_{A}^{a}}d\overline{\theta_{\bm{n}}},
\end{align}
where $\int _{\overline{E_{\xi}}}\overline{\bm{\xi}}'\cdot\overline{\bm{n}_{A}^{a}}d\overline{\theta_{\bm{n}}}=I_{2}$. Then we have
\begin{equation}
\int_{N}q_{i}(\bm{\xi})d\sigma_{\bm{\xi}}>\int_{N}q'_{i}(\bm{\xi})d\sigma_{\bm{\xi}}=\frac{\int_{\overline{E}}\overline{r(\bm{n}_{A}^{a})}d\overline{\theta_{\bm{n}}}}{I_{2}},
\end{equation}
which proves that value $\mathbb{S}_{i}$ is larger than $\mathbb{S}_{G}$. The proof of 1D ellipsoid case is similar to the 2D case. \qed\\
\section*{appendix c: Proof that translation vectors are the same in Alice to Bob case in the main text}
\indent Suppose there is a steering figure $S_{W}$ of an arbitrary Werner state in $\tilde{B}$, with geometric model $\mathfrak{G}$=$\{q_{G}(\bm{\xi}),p_{G}(a|A,\bm{\xi})\}$. And we have two geometric models $\{q(\bm{\xi}),p(a|A,\bm{\xi})\}$ and $\{q'(\bm{\xi}),p'(a|A,\bm{\xi})\}$ satisfying
\begin{align}
p(a|A,\bm{\xi})&=p'(a|A,\bm{\xi})=p_{G}(a|A,\bm{\xi})\nonumber,\\
\tilde{p}&(a|A)=\tilde{p}'(a|A),\tag{C1}
\end{align}
and
\begin{align}
&\bm{s}_{A}^{a}=\bm{s}_{W_{A}}^{a}+\bm{t}\nonumber,\\&\bm{s}_{A}^{a'}=\bm{s}_{W_{A}}^{a}+\bm{t}',\tag{C2}
\end{align}
where $\tilde{p}(a|A)=\int_{N_{C}}p(a|A,\bm{\xi})q(\bm{\xi})d\sigma$, $\bm{s}_{W_{A}}^{a}$ are the vectors that generate $S_{W}$. In this appendix we'll prove that $\bm{t}=\bm{t}'$, and therefore all g-models that realize (C1) and (C2) (corresponding to the Alice to Bob steering case in the main text) have the same translation vector $\bm{t}$.\\
\indent Let $y(\bm{\xi})$ denote the difference $q(\bm{\xi})-q(-\bm{\xi})$. First we prove that $y(\bm{\xi})=y'(\bm{\xi})$, we do it by proving that distribution $y(\bm{\xi})$ is unique for every $\{q(\bm{\xi}),p(a|A,\bm{\xi})\}$ which satisfies (C1) and (C2). We choose an arbitrary measurement $A_{0}$ and one of its outcomes $a_{0}$, let $\bm{s}_{A_{0}}^{a_{0}}$ denote its shrinked Bloch vector and $\bm{n}_{A_{0}}^{a_{0}}$ denote the outer normal vector of the steering figure corresponding to $\bm{s}_{A_{0}}^{a_{0}}$. Then we choose a set of projectors $\{P_{A'}^{a'}\}$ (we'll denote them with their Bloch vectors $\{\bm{x}_{A'}^{a'}\}$), in which every $\bm{x}_{A'}^{a'}$ has a small angle $d\theta$ between $\bm{x}_{A_{0}}^{a_{0}}$. Then we have a set of vectors $\{\bm{s}_{A'}^{a'}\}$ and corresponding outer normal vectors $\{\bm{n}_{A'}^{a'}\}$, note that each $\bm{n}_{A'}^{a'}$ also has an angle $d\theta$ between $\bm{n}_{A_{0}}^{a_{0}}$ in this case.\\
\indent Since
\begin{align}
p&(a|A,\bm{\xi})=\left \{
\begin{aligned}
1,    &\qquad \bm{\xi}\cdot\bm{n}_{A}^{a}\geq0,\\
0,    &\qquad otherwise,
\end{aligned}
\right.\nonumber\tag{C3}
\end{align}
we have
\begin{equation}
\tilde{p}(a|A)=\int_{R_{A}^{a}}q(\bm{\xi})d\sigma,\nonumber\tag{C4}
\end{equation}
where $R_{A}^{a}$ is the hemisphere consisting of unit vectors $\bm{v}$ satisfying $\bm{v}\cdot\bm{n}_{A}^{a}\geq 0$. Now we do the subtraction $\tilde{p}(a'|A')-\tilde{p}(a_{0}|A_{0})$ for all $\tilde{p}(a'|A')$, and for each $\tilde{p}(a'|A')$ there is
\begin{equation}
\tilde{p}(a'|A')-\tilde{p}(a_{0}|A_{0})=\int_{d_{A'}^{a'}}q(\bm{\xi})d\sigma-\int_{\overline{d}_{A'}^{a'}}q(\bm{\xi})d\sigma,\nonumber\tag{C5}
\end{equation}
where $d_{A'}^{a'}$ and $\overline{d}_{A'}^{a'}$ are the non-intersecting areas of $R_{A'}^{a'}$ and $R_{A_{0}}^{a_{0}}$. We plot a figure and construct a reference frame to illustrate the case more clearly. (FIG. C1) \\
\setcounter{figure}{0}
\indent Let $g(\phi)$ denote $\frac{\tilde{p}(a'|A')-\tilde{p}(a_{0}|A_{0})}{d\theta}$, where $\phi$ is one of the coordinates of $\bm{n}_{A'}^{a'}$ (see FIG. C1), and let $D_{A'}^{a'}$ denote the area $d_{A'}^{a'}\cup \overline{d_{A'}^{a'}}$. From (C5) there is
\begin{equation}
g(\phi)=\int_{D_{A'}^{a'}} q[\bm{\xi}_{(\frac{\pi}{2},\phi')}]\cos(\phi'-\phi)d\phi',\nonumber\tag{C6}
\end{equation}
where $\bm{\xi}_{(\frac{\pi}{2},\phi')}$ is the unit vector $\bm{\xi}$ with coordinates $(\frac{\pi}{2},\phi')$. $g(\phi)$ is a continuous function on $\phi\in[0,2\pi]$ determined by $\tilde{p}(a|A)$ and the choice of $\bm{x}_{A_{0}}^{a_{0}}$. For each $\phi$ there is an equation of (C6) type.\\
\indent  Let $f_{\phi}(\phi')$ denote the function
\begin{align}
f_{\phi}(\phi')=
\left \{
\begin{aligned}
&\cos(\phi-\phi'), &\quad \phi'\in[\phi-\frac{\pi}{2},\phi+\frac{\pi}{2}],\\
&0,   &otherwise,
\end{aligned}
\right.\nonumber\tag{C7}
\end{align}
then equation (C6) can also be written as
\begin{equation}
g(\phi)=\int_{0}^{2\pi} [q(\phi')-q(-\phi')]f_{\phi}(\phi')d\phi',\nonumber\tag{C8}
\end{equation}
where $q(\bm{\xi})$ is written as $q(\phi')$ since each $\bm{\xi}$ can be represented by $(\frac{\pi}{2},\phi')$ in the reference frame. Here we would also write $y(\bm{\xi}')$ as $y(\phi')$, denoting the term $q(\phi')-q(-\phi')$.\\
\begin{figure}
\centering
\includegraphics[width=1.8in,height=1.8in]{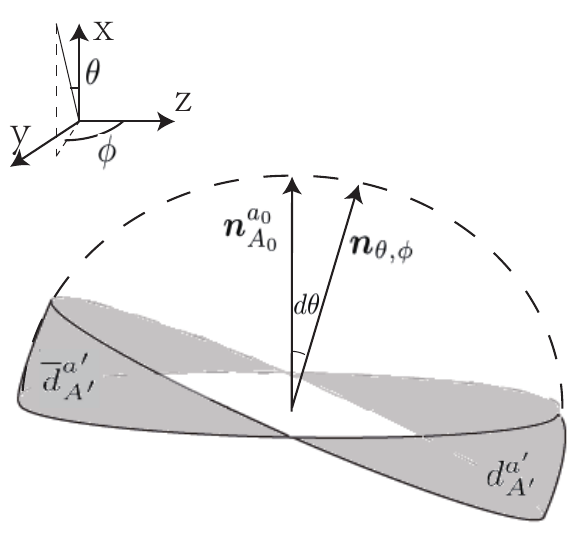}
\renewcommand{\figurename}{FIG.C}\\
\caption{The shadowed (grey) areas are $d_{A'}^{a'}$ and $\overline{d}_{A'}^{a'}$. As we have constructed the reference frame based on $\bm{n}_{A_{0}}^{a_{0}}$, all $\bm{n}_{A'}^{a'}$ can be written $\bm{n}_{(\theta,\phi)}$ of some $\theta,\phi$.}
\label{fig:graph}
\end{figure}
\indent Let $T(l)$ denote the translation operator in the circle area consisting of $\bm{\xi}$ with $\theta=\frac{\pi}{2}$, satisfying $T(l)f(\phi)=f(\phi-l)$ for any functions on the circle. The set of all functions $\{f_{\phi}(\phi')|\phi\in[0,2\pi)\}$ can be generated by performing $T(l')$ with $l'\in [0,2\pi)$ on one arbitrary function in the set. The eigenfunctions of $T(l')$ in the circle area $\phi'\in[0,2\pi]$ are $\{\exp (in\phi')|n\in Z\}$ with the eigenvalues $\exp(-inl')$, where $Z$ is the set of integers. This set of eigenvectors is also a set of basis of the circle area. \\
\indent Now we choose an $f_{\phi_{0}}(\phi')$ where $\phi_{0}=0$ and expand it using the basis, that is $f_{\phi_{0}}(\phi')=\sum_{n} F_{\phi_{0}}(n)\exp({in\phi'})$, then other $f_{\phi}(\phi')$ can be written
\begin{equation}
f_{\phi}(\phi')=T(l')f_{\phi_{0}}(\phi')=\sum_{n}\exp{(-inl')}F_{\phi_{0}}(n)\exp({in\phi'}),\nonumber\tag{C9}
\end{equation}
where $l'=\phi-\phi_{0}=\phi$ here.\\
\indent Function $g(\phi)$ can also be expanded as $g(\phi)=\sum_{n} G(n)\exp({in\phi})$, combining (C8) and (C9) we have
\begin{equation}
G(n)=F_{\phi_{0}}(n) \int_{0}^{2\pi} y(\phi')\exp{(in\phi')}d\phi'.\nonumber\tag{C10}
\end{equation}
where the integral part is just $Y(-n)$, the expansion coefficient of $y(\phi')$ under the basis (constant factor ignored). So there is $Y(n)=k\cdot\frac {G(-n)}{F_{\phi_{0}}(-n)}$ ($k$ is a constant), which shows that if solution $y(\phi)$ ($Y(n)$) exists for a $g(\phi)$, it's an unique one. By above process we have proved that $y(\bm{\xi})$ is unique under (C1) on area $\theta=\frac{\pi}{2}$. Since $A_{0}$ and $a_{0}$ are arbitrarily chosen, we can repeat the process for all projectors and prove that $y(\bm{\xi})$ is unique on the whole surface of sphere $\tilde{B}$. \\
\indent Then we prove that $\bm{t}=\bm{t}'$ in such case. We choose an arbitrary measurement $A_{1}$, and denote its two outcomes with $a_{+}$ and $a_{-}$. Substituting Eqs. (5) and (6) in the main text into Eqs. (C1) and (C2), after simple calculations we can obtain that
\begin{align}
\bm{t}-\bm{t}'=\int_{R_{+}} [q(\bm{\xi})-q'(\bm{\xi})]\cdot\bm{\xi}d\sigma,\tag{C11}
\end{align}
where $R_{+}$ is the hemisphere satisfying $p(a_{+}|A_{1},\bm{\xi})=1$. We can also obtain that
\begin{align}
\bm{t}-\bm{t}'=\int_{R_{-}} [q(\bm{\xi})-q'(\bm{\xi})]\cdot\bm{\xi}d\sigma,\tag{C12}
\end{align}
where $R_{-}$ is the hemisphere satisfying $p(a_{-}|A_{1},\bm{\xi})=1$. Using former result we have
\begin{align}
q(\bm{\xi})-q(-\bm{\xi})=q'(\bm{\xi})-q'(-\bm{\xi})=y(\bm{\xi}),\tag{C13}
\end{align}
so we can change the integrating area of (C12) into $R_{+}$, and (C12) becomes
\begin{align}
\bm{t}-\bm{t}'=\int_{R_{+}} [q(\bm{\xi})-q'(\bm{\xi})]\cdot(-\bm{\xi})d\sigma.\tag{C14}
\end{align}
\indent Comparing (C11) and (C14) we have $\bm{t}-\bm{t}'=\bm{0}$, which implies that any models which realize probability conditions (C1) and (C2) have the same translation vectors $\bm{t}$.

\section*{appendix d: Proof of the steering criterion for two-qudit states}
\renewcommand{\theequation}{D\arabic{equation}}
\setcounter{equation}{0}
First we restate the criterion formally as follow:\\
\indent$\bf{Criterion.}$ An measurement assemblage of two-qudit quantum state admits an LHS model if and only if $\mathbb{S}_{o}\leq1$ for its steering figure.\\
\noindent \emph{Proof}.
For the necessity, since every qudit quantum state can be written as mixture of qudit pure states, any LHS model can be rewritten as $\{p(a|A,\lambda),q(\lambda),\rho(\lambda)\}$, where $\rho(\lambda)=\frac{1}{d}(\bm{I}+\bm{b}_{\lambda}\cdot \bm{\gamma})$ is qudit pure state, $\bm{b}_{\lambda}$ is the Bloch vector determined by $\lambda$. From definition we know that the measurement assemblage $\{\tilde{\rho}_{A}^{a}\}$ can be realized by LHS model $\{p(a|A,\lambda),q(\lambda),\rho(\lambda)\}$ using Eq. (1) in the main text. Let
\begin{align}
q(\bm{\eta})&=\sum_{\{\lambda|\bm{b}_{\lambda}=\bm{\eta}\}}q(\lambda)\nonumber,\\
p(a|A,\bm{\eta})=&\frac{\sum_{\{\lambda|\bm{b}_{\lambda}=\bm{\eta}\}}p(a|A,\lambda)q(\lambda)}{q(\bm{\eta})},
\end{align}
then Eqs. (36) and (37) would hold for $\bm{s}_{A}^{a}$ of all $\tilde{\rho}_{A}^{a}$, therefore $\{q(\bm{\eta})$, $p(a|A,\bm{\eta})\}$ is a geometric model for the steering figure. Now that $\mathbb{S}=\int_{\Lambda}q(\bm{\eta})d\bm{\eta}=\int q(\lambda)d\lambda=1$, $\mathbb{S}_{o}\leq \mathbb{S}=1$.\\
\indent For the sufficiency, let $p_{o}(a|A,\bm{\eta})$ and $q_{o}(\bm{\eta})$ denote the distributions of the optimal geometric model for the steering figure. When $\mathbb{S}_{o}\leq 1$, we can construct the modified distributions $\{q_{o}'(\bm{\eta}),p_{o}'(a|A,\bm{\eta})\}$ satisfying
\begin{align}
&q_{o}'(\bm{\eta})=(1-\mathbb{S}_{o})\cdot\delta(\bm{\eta})+q_{o}(\bm{\eta}),\nonumber\\
p_{o}'(a|A,&\bm{\eta})=\left \{
\begin{aligned}
&\frac{p(a|A)-I(a|A)}{\int_{C}q_{o}'(\bm{\eta})d\bm{\eta}}, &\quad \bm{\eta}=\bm{0},\\
&p_{o}(a|A,\bm{\eta}), &\quad otherwise,
\end{aligned}
\right.
\end{align}
 where $\delta(\bm{\eta})$ is Dirac delta function, $I(a|A)$ denotes the integral $\int_{\Lambda} p_{o}(a|A,\bm{\eta})q_{o}(\bm{\eta})d\bm{\eta}$, and the subscript $C$ in the integral expressions means that the integral area is a small neighbourhood of the center $C$. Now we have $\int_{\Lambda_{C}}q'_{o}(\bm{\eta})d\bm{\eta}=1$. Then we can construct a model $\{p(a|A,\bm{\eta}),\rho(\bm{\eta}),q(\bm{\eta})\}$ by  $\rho(\bm{\eta})=\frac{1}{d}(\bm{I}+\bm{\eta}\cdot\bm{\gamma})$, $q(\bm{\eta})=q_{o}'(\bm{\eta})$, $p(a|A,\bm{\eta})=p_{o}'(a|A,\bm{\eta})$. Note that if $\int_{C}q_{o}'(\bm{0})d\bm{\eta}=0$ in Eq. (D2), we let $p_{o}'(a|A,\bm{0})=\frac{1}{d}$. This model satisfies equations (1) (2), thus is an LHS model for the state.\qed\\

\section*{appendix e: Proof of the extension for two-qudit isotropic states}
\renewcommand{\theequation}{E\arabic{equation}}
\setcounter{equation}{0}

\emph{Proof.}  Using the symmetry, we can let distribution $q_{G}(\bm{\eta})$ be a constant independent of $\bm{\eta}$, so model $\{p_{G}(a|A,\bm{\eta}),q_{G}(\bm{\eta})\}$ satisfies the conditions in Eqs. (36). By projecting both sides of equation (37) onto the corresponding $\bm{n}_{A}^{a}$, a new equation is obtained
\begin{equation}
r(\bm{n}_{A}^{a})=\int _{\Lambda} p(a|A,\bm{\eta})q(\bm{\eta})\bm{\eta}\cdot\bm{n}_{A}^{a} d\bm{n},
\end{equation}
where $r(\bm{n}_{A}^{a})=\bm{s}_{A}^{a}\cdot\bm{n}_{A}^{a}$.\\
\indent Then, integrating both sides of Eq. (E1) with respect to $\bm{n}_{A}^{a}$ over $\Omega$, we have
\begin{equation}
\int_{\Omega}r(\bm{n}_{A}^{a})d\bm{n}=\int_{\Lambda}q(\bm{\eta})\int _{\Omega}p(a|A,\bm{\eta}) \bm{\eta}\cdot\bm{n}_{A}^{a}d\bm{n}d\bm{\eta},
\end{equation}
substituting model $\{q_{G}(\bm{\eta}),p_{G}(a|A,\bm{\eta})\}$ into Eq.(E2), we have
\begin{equation}
\int_{\Omega}r(\bm{n}_{A}^{a})d\bm{n}=q_{G}(\bm{\eta})\int_{\Lambda}\int _{N_{\bm{\eta}}} \bm{\eta}\cdot\bm{n}_{A}^{a}d\bm{n}d\bm{\eta},
\end{equation}
where $N_{\bm{\eta}}$ is the region consists of unit vectors $\bm{n}_{A}^{a}$ whose corresponding region $L_{A}^{a}$ contains $\bm{\eta}$ ($\bm{\eta}\in L_{A}^{a}$).\\
\indent Let $k_{d}$ denote the integral $\int _{N_{\eta}} \bm{\eta}\cdot\bm{n}_{A}^{a}d\bm{n}$. For two-qudit isotropic states, $\bm{n}_{A}^{a}$ is parallel to $\bm{s}_{A}^{a}$, hence $\bm{n}_{A}^{a}$ is one to one proportional to some $\bm{\eta}'$ with relation $\bm{n}_{A}^{a}=\frac{1}{|\bm{s}_{d}|}\bm{\eta}'$, and region $\Omega$ is proportional to region $\Lambda$ with a factor $\frac{1}{|\bm{s}_{d}|}$. Consider the symmetry of vectors $\bm{\eta}$ in $\Lambda$ , $k_{d}$ is independent of $\bm{\eta}$. Then equation (E3) becomes
\begin{equation}
\int_{\Omega}r(\bm{n}_{A}^{a})d\bm{n}=k_{d}\cdot  \int_{\Lambda}q_{G}(\bm{\eta})d\bm{\eta}=k_{d}\cdot \mathbb{S}_{G}.
\end{equation}
\indent From (E4) we obtain
\begin{equation}
\mathbb{S}_{G}=\frac{\int_{\Omega}r(\bm{n}_{A}^{a})d\bm{n}}{k_{d}}.
\end{equation}
\indent Consider another geometric model $\mathcal{G}_x$. Using equation (E2) we have
\begin{align}
\int_{\Omega}r(\bm{n}_{A}^{a})d\bm{n}&=\int_{\Lambda}q_{x}(\bm{\eta})\int _{\Omega} p_{x}(a_{i}|A,\bm{\eta})\bm{\eta}\cdot\bm{n}_{A}^{a} d\bm{n} d\bm{\eta}\nonumber\\ &\leq \int_{\Lambda}q_{x}(\bm{\eta})\int _{N_{\bm{\eta}}} \bm{\eta}\cdot\bm{n}_{A}^{a_{0}} d\bm{n} d\bm{\eta}\nonumber\\
&= k_{d} \cdot \int_{\Lambda}q_{x}(\bm{\eta})d\bm{\eta}.
\end{align}
The inequality in (E6) comes as $\sum_{i=0}^{d-1} p_{x}(a_{i}|A,\bm{\eta})\bm{\eta}\cdot\bm{n}_{A}^{a_{i}}\leq \bm{\eta}\cdot\bm{n}_{A}^{a_{0}}$, where $\{a_{i}\}$ are the outcomes of measurement A, $a_{0}$ is the outcome whose corresponding region $L_{A}^{a_{0}}$ contains $\bm{\eta}$ $(\bm{\eta}\in L_{A}^{a_{0}})$.  Since $\mathbb{S}_{x}=\int_{\Lambda_{C}}q_{x}(\bm{\eta})d\bm{\eta}=p(\bm{0})+\int_{\Lambda}q_{x}(\bm{\eta})d\bm{\eta}$, inequality (E6) indicates that
$\mathbb{S}_{x}\geq\frac{\int_{\Lambda}r(\bm{n}_{A}^{a})d\bm{n}}{k_{d}}=\mathbb{S}_{G}$, thus the extension is proved correct for two-qudit isotropic states.\qed

\end{document}